\documentclass[preprint,showpacs,preprintnumbers,amsmath,amssymb]{revtex4}

\usepackage{graphicx}% Include figure files
\usepackage{dcolumn}% Align table columns on decimal point
\usepackage{bm}% bold math
\usepackage{color}
\usepackage{subfigure}
\usepackage{epsfig}
\usepackage{amsmath}
\begin{document}

\preprint{}

\title{Strangeness content of the nucleon in quasielastic
  neutrino-nucleus reactions} %

\author{N.~Jachowicz}
\author{P.~Vancraeyveld}
\author{P.~Lava}
\author{C.~Praet}
\author{J.~Ryckebusch}

\affiliation{Department of Subatomic and Radiation Physics, Ghent
  University,
  Proeftuinstraat 86, B-9000 Gent, Belgium}%Lines break automatically or can be forced with \\

\date{\today}% It is always \today, today,

\begin{abstract}
We present a systematic study of  the sensitivity of quasielastic neutrino-nucleus cross sections at intermediate energies to the strange quark sea of the nucleon. To this end, we investigate  the impact of the weak strangeness form factors on the ratio of proton-to-neutron knockout, the ratio of neutral-to-charged current cross sections, on the Paschos-Wolfenstein relation, and on the longitudinal helicity asymmetry. The influence of axial as well as vector strangeness effects is discussed.  For the latter, we introduce strangeness parameters from various hadron models and from a recent fit to data from parity violating electron scattering.

In our  model, the nuclear target is described in terms of a relativistic mean-field approach.  The effects of  final-state interactions on the outgoing nucleon are quantified within a relativistic multiple-scattering Glauber approach.  Our results are illustrated with cross sections for the scattering of 1 GeV neutrinos and antineutrinos off a $^{12}$C target.

  Folding with a proposed FINeSSE (anti)neutrino energy-distribution has no qualitative influence on the overall sensitivity of the cross-section ratios to strangeness mechanisms. We show that vector strangeness effects are large and strongly $Q^2$ dependent. 
\end{abstract} 

\pacs{25.30.Pt; 24.10.Jv; 24.70.+s; 14.20.Dh}% PACS, the Physics and Astronomy
\maketitle

\section{Introduction}
In recent years, the role of the strangeness content of the nucleon has been the subject of  numerous experimental and theoretical studies.  These efforts aim at  determining  the role of the strange quark sea in the nucleon's axial current and in its charge and magnetization distributions.

 Measurements of the strange vector form factors in parity violating electron scattering  (PVES) point to  small strangeness form factors \cite{Liu,beise}.  
In a number of experiments around $Q^2$=0.1 (GeV/c)$^2$ \cite{happex1,happex2,happex3}  and at $Q^2$=0.48 (GeV/c)$^2$ \cite{happex4,happex5}, the HAPPEX collaboration  measured the strange electric form factor $G_E^s$.  The data produce $G_E^s$ and $G_M^s$ values consistent with zero at low $Q^2$, hence  a hadron model with no strangeness contributions is  compatible with the data.  
The PVA4  experiment \cite{PVA4l,PVA4h} measured the parity violating asymmetry in polarized electron elastic scattering at  $Q^2$=0.1 and 0.230 (GeV/c)$^2$ and forward electron scattering angles, and found a negative asymmetry that is smaller than the one expected in the absence of strangeness contributions.  They extracted a combination  $G_E^s+\eta G_M^s$ that is positive, indicating small values for the strangeness form factors, but consistent with zero \cite{PVA4h,PVA4l}.  A combination of these results with those of the SAMPLE experiment at MIT-Bates \cite{sample1,sample2}, that isolated the magnetic contribution $G_M^s$ in a low-energy measurement at backward scattering angles for $Q^2$ values of 0.1 and 0.04 (GeV/c)$^2$, hints at small and positive values for $G_E^s$ and $G_M^s$ \cite{PVA4h}.  The Jefferson Lab G0 \cite{G0} experiment made measurements over the broad $Q^2$ range 0.12 (GeV/c)$^2\leq Q^2 \leq$ 1.0 (GeV/c)$^2$ 
hinting at a non-zero strangeness signal with a 89\% confidence limit.  The $Q^2$ dependence of the extracted $G_E^s+\eta G_M^s$ is quite remarkable.  It might indicate negative values for $G_E^s(Q^2)$
 for $Q^2$ values up to 0.3 (GeV/c)$^2$, and shows a clear tendency towards positive values of $G_E^s+\eta G_M^s$ at larger $Q^2$.  These results are consistent with the HAPPEX, SAMPLE and PVA4 ones.  

Although the value of the strangeness contribution to the axial current and its $Q^2$ dependence is not precisely known yet, it is clear that strangeness has a considerable impact on the axial form factor.  Polarized lepton deep inelastic scattering experiments carried out by the SMC \cite{SMC} and HERMES \cite{HERMES} collaborations pointed to negative values for $\Delta s$ of the order $\Delta s \approx$-0.1.  Unfortunately, quite some uncertainty stems from the fact that the extraction of the first moment of the strange quarks' helicity distribution from the polarized structure function is subject to assumptions concerning the validity of $SU(3)$ flavor symmetry in hyperon $\beta$-decays and extrapolation of the spin structure function to vanishing 
Bjorken $x$.  In PVES the heavily suppressed axial strangeness contribution is veiled by radiative corrections. These are not present in weak processes.

Neutrino scattering experiments are considered to be the optimal tool for extracting information about the contribution of strange quarks to the axial current.  As in electromagnetic processes, ratios are often used to enhance the sensitivity of experimental data to strangeness parameters and reduce uncertainties related to nuclear effects.
An important effort in this direction was the Brookhaven E734 experiment \cite{bnl734} that measured neutrino and antineutrino scattering off protons in the range   0.4 (GeV/c)$^2 \,\leq\, Q^2\,\leq\, 1.1$ (GeV/c)$^2$, and used the ratio of neutral-to-charged current cross sections to extract information on the strange spin of the proton. Unfortunately, its results suffer strongly from experimental uncertainties.
More recent reanalyses \cite{pate,albe,garvey} point to $\Delta s$ values of approximately $\Delta s\approx-0.21$, but the systematic errors on the data are too large to provide conclusive results.
The proposed FINeSSE experiment plans to improve on this by measuring the neutral-to-charged current ratio in the range   0.2 (GeV/c)$^2\,\leq\, Q^2 \,\leq$ 1.0 (GeV/c)$^2$ \cite{finesse} with higher statistics and substantially reduced errors. 
The  MINER$\nu$A experiment aims at high-precision measurements of neutrino scattering cross sections, and would be well-suited to examine the $Q^2$ evolution of the strangeness form factors  in quasielastic scattering at $Q^2\geq$ 1(GeV/c)$^2$ \cite{minerva}.  BooNE may be able to contribute to this work through an analysis of the 
neutrino-minus-antineutrino neutral-over-charged current cross-section ratio \cite{boone}.

As pointed out in Ref.~\cite{pate}, one of the major sources of uncertainties  in strangeness studies is the mutual influence of  vector  and axial strangeness  on each other, in their effect on  cross sections and cross-section ratios.  

Hence, a combined analysis of the PVES and neutrino-scattering data  is a prerequisite for a thorough understanding of the proton's strangeness properties \cite{pate}.

Hadron models have made widely different predictions for the vector strangeness quantities \cite{jaffe89,musolf94,kim95,silva01,cbm,hybrid,chiralquark,disprel,leinweber,lewis,lyubo}.
Several recent theoretical studies of neutrino-nucleus scattering at intermediate energies \cite{vanderventel,meucci,praet,lava} highlighted particular aspects of the strangeness glimpses that can be caught in quasielastic neutrino-scattering experiments. In this paper, we aim at providing a more systematic overview of the sensitivity of various neutrino cross-section ratios to the nucleon's strange quark content. Next to the well-studied ratios of proton-over-neutron knockout and neutral-over-charged current cross sections, we  look into the influence of  strangeness on the Paschos-Wolfenstein relation and the longitudinal helicity asymmetry. 
We study the strangeness effects on these ratios as a function of the kinetic energy of the ejected nucleon, and   pay attention to the $Q^2$ dependence of the strangeness impact.
As the influence of the vector and the axial strangeness on neutrino cross-section ratios is strongly intertwined, we compared the $Q^2$ dependent behavior of these ratios for different hadron models for the vector strangeness parameters  and for a fit to the G0 data \cite{Liu}.  Strong effects from the vector strangeness make the extraction of
the axial strangeness properties of the nucleon from ratios of
neutrino-nucleus responses a challenging task.

In the following two sections, we present the formalism used to describe the 
neutrino scattering processes and the parameterization of the nucleon's strange quark sea.
Section \ref{secres} illustrates the influence of strangeness on different cross-section ratios for 1 GeV neutrino and antineutrino scattering off $^{12}$C.  
Finally, in Section \ref{secfin} we show some results for FINeSSE, and discuss the sensitivity of the ratios to the strangeness parameters as a function of four-momentum transfer in section \ref{conclu}.

\section{Neutrino-nucleus scattering at intermediate energies}\label{secform}
A full derivation of (anti)neutrino-induced one-nucleon knockout cross sections can for example be found in  Ref.~\cite{us}.  The one-fold differential cross section  is given by
\begin{equation}
\frac{d\sigma}{dT_N}=\frac{M_N M_{A-1}}{(2\pi)^3 M_A} 4\pi^2 \int \sin\theta_ld\theta_l\int\sin\theta_Nd\theta_N k_N f_{rec}^{-1}\sigma_M\left[v_LR_L+v_TR_T+hv_{T'}R_{T'}\right],\label{cs1}
\end{equation}
with $M_N$, $T_N$ and $\vec{k}_N$ the mass, kinetic energy and momentum of the ejectile, $M_A$ and $M_{A-1}$ the mass of the target and residual nuclei.  The direction of the outgoing  lepton and nucleon is determined by $\Omega_l(\theta_l,\phi_l)$ and $\Omega_N(\theta_N,\phi_N)$. The recoil factor is denoted by $f_{rec}$. The quantity $\sigma_M$ is the weak variant of the Mott cross section
\begin{equation}
\sigma_M^Z =
  \left(\frac{G_F\cos(\theta_l/2)\varepsilon'M_Z^2}{\sqrt{2}\pi(Q^2+M_Z^2)}\right)^2,
\end{equation}
for neutral current processes, and 
\begin{equation}
\sigma_M^{W^{\pm}} =\sqrt{1-{\frac{M_l^{'2}}{\varepsilon'^2}}}
  \left(\frac{G_F\cos\theta_c\cos(\theta_l/2)\varepsilon'M_W^2}{{2}\pi(Q^2+M_W^2)}\right)^2,
\end{equation}
for charged current reaction.  In these equations, $G_F$ is the weak interaction Fermi coupling constant, $\theta_c$ the Cabibbo angle, $M_Z$ and $M_W$ the weak boson masses, $M_l'$ the mass of the outgoing lepton, $\varepsilon'$ the energy of the outgoing lepton, and $Q^2=-q_{\mu}q^{\mu}$ the transferred four momentum.  
In Eq.(~\ref{cs1}), $v_L$, $v_T$ and $v_{T'}$ are the longitudinal, transverse and interference kinematic factors and $R_L$, $R_T$ and $R_{T'}$ the accompanying structure functions, reflecting the influence of nuclear dynamics on the scattering process \cite{us}. The helicity of the incoming neutrino is denoted by $h$.
The basic quantities in the computation of these response functions are the transition matrix elements $\langle J^{\mu}\rangle$.  Within an independent-nucleon model and adopting the impulse approximation, the matrix elements of the weak current operator $\widehat{J}^{\mu}$ can be expressed as
\begin{equation}
\langle J^{\mu}\rangle=\int d\vec{r}\; \overline{\phi}_F(\vec{r})\widehat{J}^{\mu}(\vec{r})e^{i\vec{q}\cdot\vec{r}}\phi_B(\vec{r}),\label{matel}
\end{equation}
with ${\phi}_B(\vec{r})$ and ${\phi}_F(\vec{r})$  the relativistic bound-state and scattering wave functions, respectively.  In our numerical calculations, bound-state wave-functions are obtained within the Hartree approximation to the $\sigma-\omega$ model \cite{serot}, adopting the W1 parameterization for the different field strengths \cite{w1}.

The influence of final-state interactions (FSI) on the ejected nucleon is examined adopting a relativistic multiple-scattering Glauber approximation (RMSGA) \cite{jan,us}. As a semi-classical approach, this technique exploits the advantages of the kinematics conditions reigning at sufficiently high energies, where high momentum transfers strongly favor forward elastic scattering of the outgoing nucleon.  The Glauber technique assumes linear trajectories for the ejectile and frozen spectator nucleons in the residual system.  The influence of the nuclear medium on the outgoing nucleon's wave function are condensed in the eikonal phase ${\cal G}[\vec{b}(x,y),z]$, that summarizes the effects of the scattering reactions the ejectile undergoes.  This results in a scattering wave function that can be written as
\begin{equation}
{\phi}_F(\vec{r})={\cal G}[\vec{b}(x,y),z]\;\phi_{k_N,s_N}(\vec{r}),
\end{equation}
with $\phi_{k_N,s_N}(\vec{r})$ a relativistic plane wave.  In the limit of vanishing final-state interactions (${\cal G}=1$) , the formalism becomes equivalent to the relativistic plane wave impulse approximation (RPWIA) \cite{us,lava}.

\section{Strangeness in the nucleon and the weak interaction}\label{strangeform}

The weak one-body vertex function related to the nuclear currents of Eq.~(\ref{matel}) is used in its cc2 form~:
\begin{eqnarray}
\label{eq:cc2ch5}
\widehat{J}^{\mu}_{cc2} &=& {F}_1^Z(Q^2)\gamma^{\mu} +
i\frac{\kappa}{2M_N}{F}_2^Z(Q^2)\sigma^{\mu\beta}q_{\beta}  +
G_A(Q^2)\gamma^{\mu}\gamma_5 + \frac{1}{2M_N} G_P(Q^2)q^{\mu}\gamma_5\; ,
\end{eqnarray}
where ${F}_1^Z$ is the weak Dirac, ${F}_2^Z$ the weak Pauli,
$G_A$ the axial and $G_P$ the pseudoscalar form factor.
Strangeness contributes to the form factors for weak neutral-current processes.
The axial form factor then becomes
\begin{equation}
G_A(Q^2)=\frac{1}{2}\frac{-g_A\tau_3+g_A^s}{(1+\frac{Q^2}{M_A^2})^2},
\end{equation}
with $g_A$=1.262, and $M_A=1.032$ GeV.
The isospin operator $\tau_3$ equals +1 for protons, -1 for neutrons. 

\begin{figure}[htb]
%figuur1
\vspace{13.8 cm}
\begin{center}
\special{hscale=30 vscale=30 hsize=1500 vsize=600
         hoffset=0 voffset=435 angle=-90 psfile="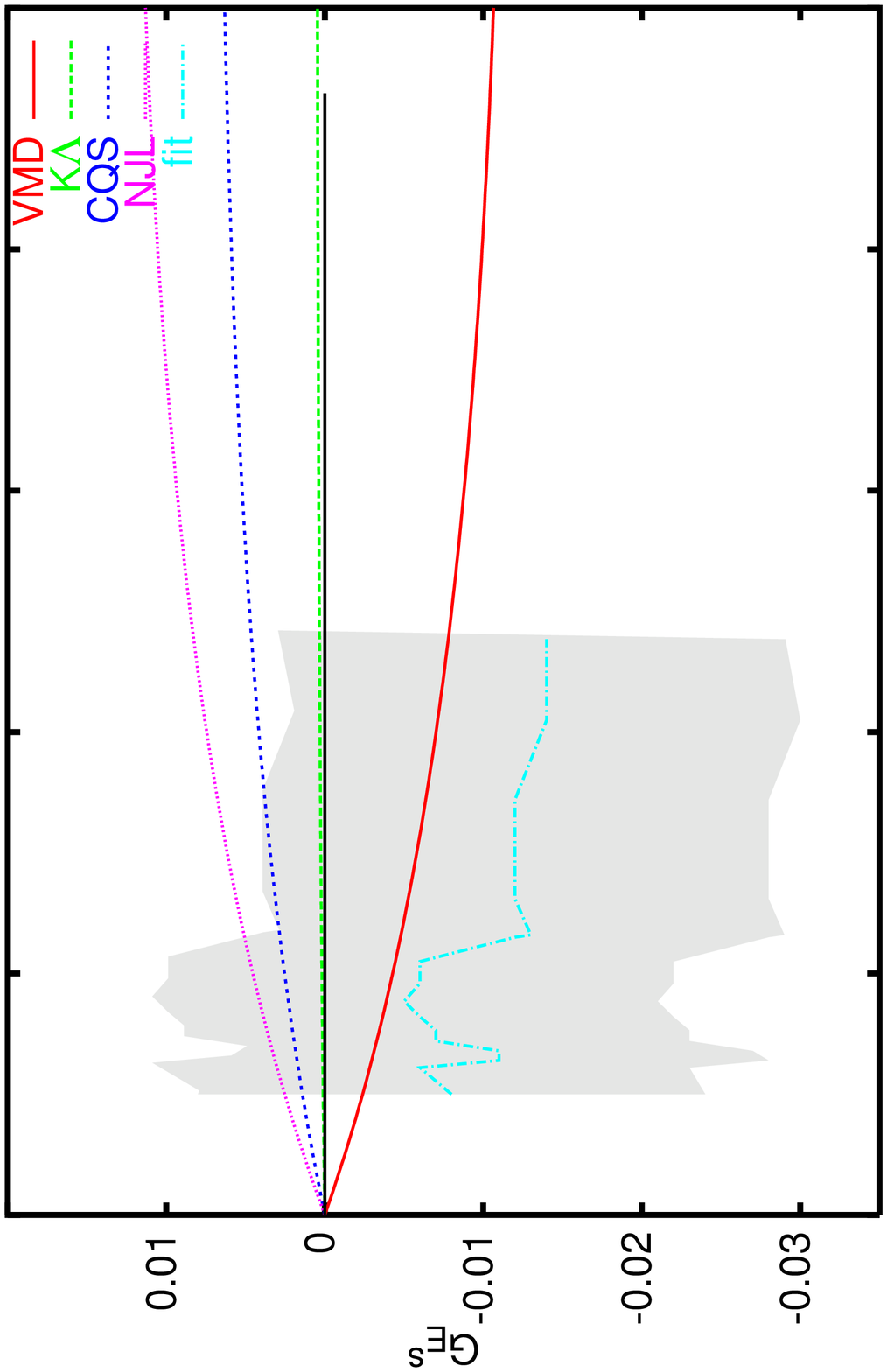"}   
\special{hscale=29.6 vscale=30 hsize=1500 vsize=600
         hoffset=2.6 voffset=302 angle=-90 psfile="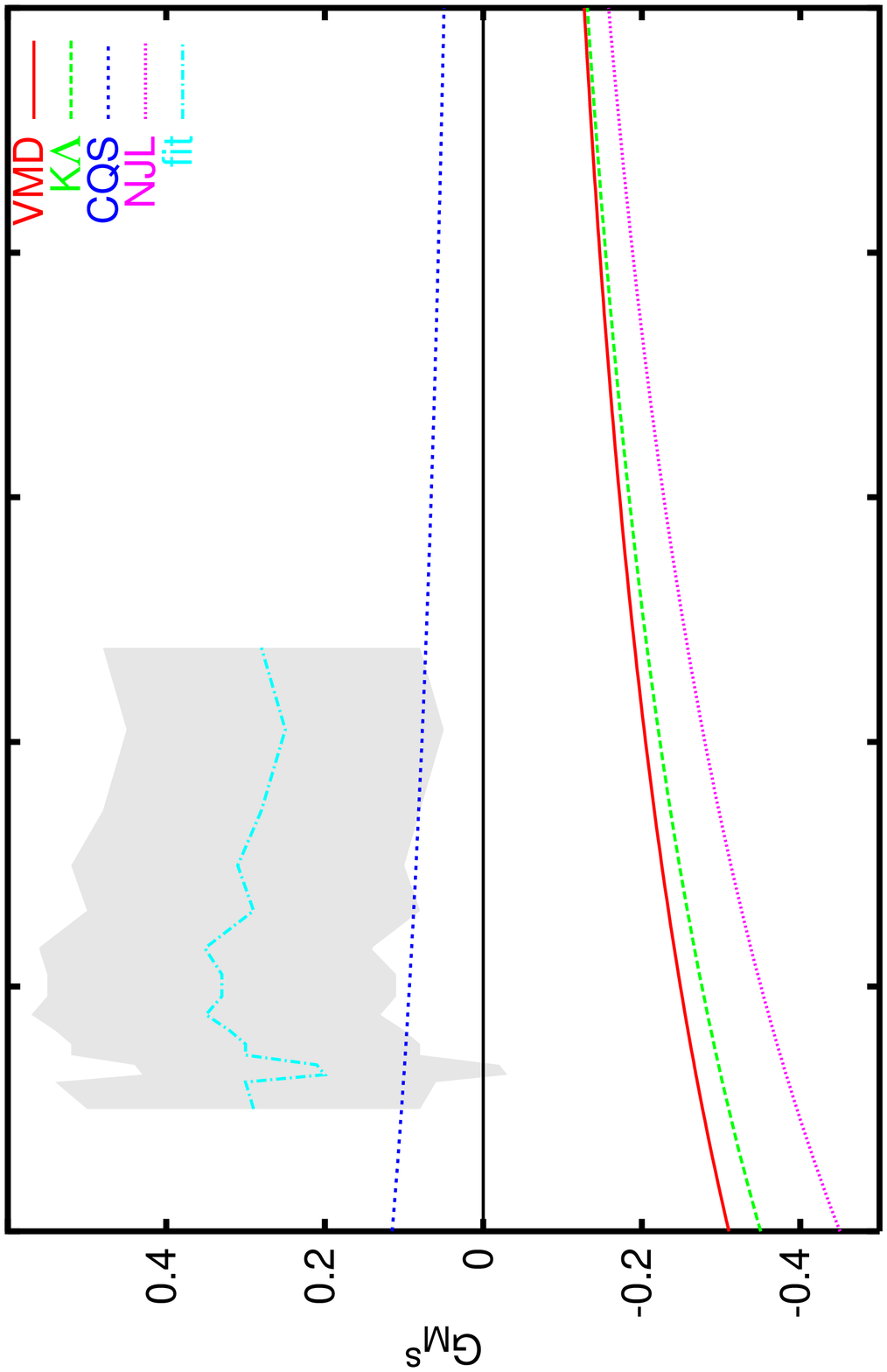"}   
\special{hscale=30 vscale=30 hsize=1500 vsize=600
         hoffset=0 voffset=168 angle=-90 psfile="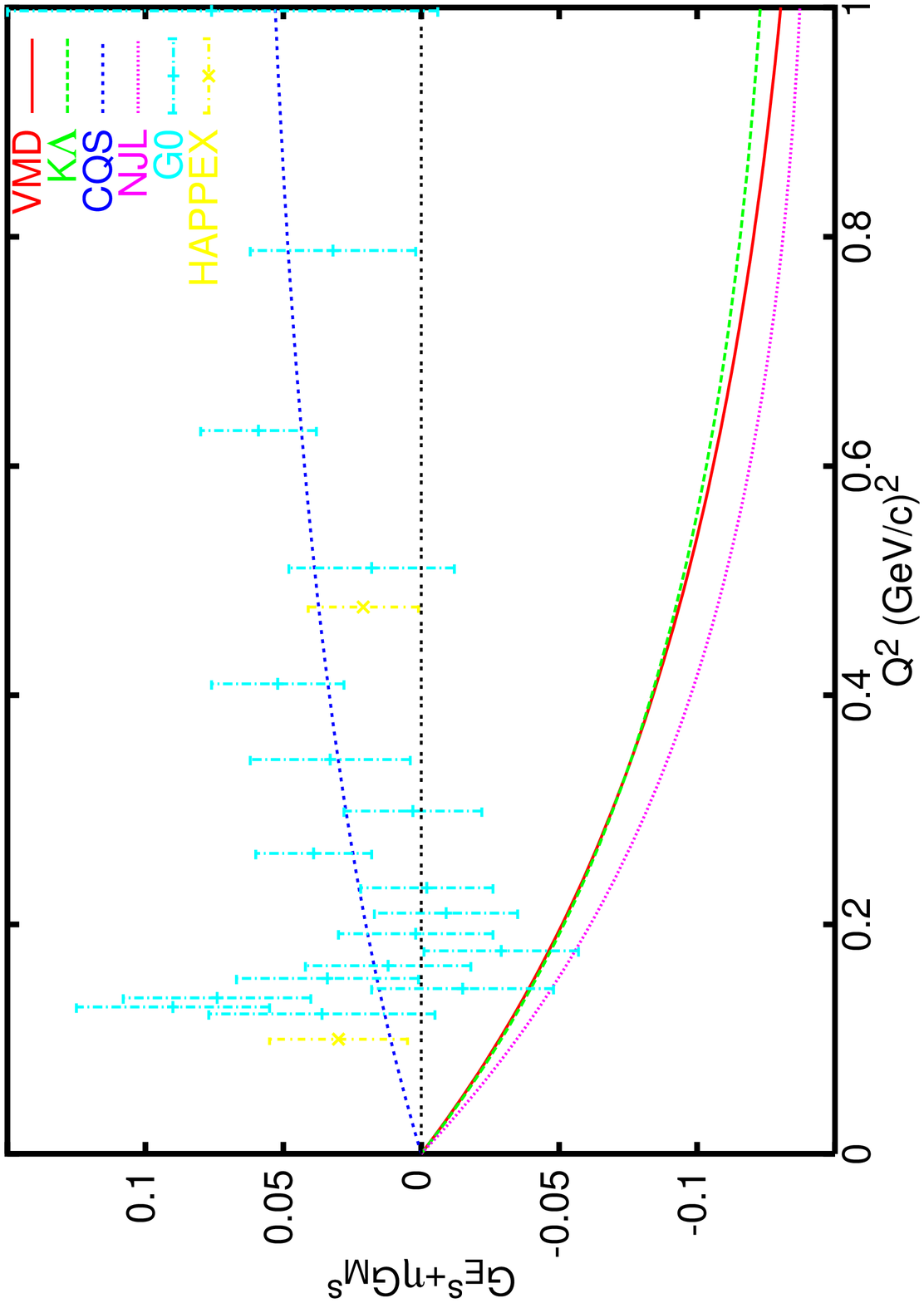"}   
    
\caption{(Color online) $Q^2$ evolution of the strange proton electric (upper panel) and magnetic
    (middle panel)  vector form factors.  The full, dashed, short-dashed and dotted  curves include  strangeness
    contributions in the parameterization of Eqs.~(\ref{eq:f1forkel})
    and (\ref{eq:f2forkel}). The adopted values for $r_s^2$ and
    $\mu_s$ are those of four different hadron models (VMD
    \cite{jaffe89}, K$\Lambda$ \cite{musolf94}, NJL \cite{kim95} and
    CQS(K)  \cite{silva01}) and can be found in
    Table~\ref{tab}.  The dashed-dotted curve represents the fit to the G0 data of Ref.~\cite{Liu}, the 1$\sigma$ error bars \cite{Liu} are indicated by the shaded region.
The lower panel shows the  combination $G_E^s+\eta G_M^s$  for different hadron models and experimental data.}
    \label{ffs}
  \end{center}
\end{figure}

The neutral-current vector form-factors are parameterized as
\begin{equation}
F_i^Z=\left(\frac{1}{2}-\sin^2\theta_W\right)\left(F_{i,p}^{EM}-F_{i,n}^{EM}\right)\tau_3-\sin^2\theta_W\left(F_{i,p}^{EM}+F_{i,n}^{EM}\right)-\frac{1}{2}F_i^s \;\;\; (i=1,2),
\end{equation}
in terms of the electromagnetic form factors  $F_{i,N}^{EM}$, with $\sin^2\theta_W=0.2224$. The $Q^2$ dependence of the vector form factors is established by means of a standard dipole parameterization. $F_i^s$ represents the strangeness contribution to the vector form factors.  We adopt the parameterization based on the three-pole ansatz of Forkel {\em et al.} \cite{forkel}
\begin{eqnarray}
\label{eq:f1forkel}
F_1^s &=& \frac{1}{6} \frac{-r_s^2Q^2}{(1 + Q^2/M_1^2)^2} \; ,\\
\label{eq:f2forkel}
F_2^s &=& \frac{\mu_s}{(1 + Q^2/M_2^2)^2} \; ,
\end{eqnarray}
with cut-off parameters $M_1 \mbox{=} 1.30$ GeV and $M_2 \mbox{=} 1.26$ GeV. The strangeness parameters $r_s^2$ and $\mu _s$ predicted by various hadron models are summarized in Fig.~\ref{ffs} and Table~\ref{tab}.
The list is incomplete, more studies can for example be found in Refs.~\cite{cbm,hybrid,chiralquark,disprel,leinweber,lewis,lyubo}.  The presented models were  selected so as to span the whole range of values in the predictions for $r_s^2$ and $\mu_s$.  
Contrary to experimental results hinting at small positive $\mu_s$, the model predictions exhibit a tendency towards a mildly negative strangeness magnetic moment and a small negative strangeness radius.  As Fig.~\ref{ffs}  illustrates, the effect of strangeness  on the electric and magnetic Sachs form factors
\begin{eqnarray}
G_E^s(Q^2)&=&F_1^s(Q^2)-\frac{Q^2}{4M_N^2}F_2^s(Q^2), \\
G_M^s(Q^2)&=&F_1^s(Q^2)+F_2^s(Q^2),
\end{eqnarray}
can be considerable. The model predictions are compared with the fit to the G0 data of Ref.~\cite{Liu}.      Apart from the $G_E^s$ value of the VMD model and the $G_M^s$ value of the CQS model, their signs differ from the ones the data are hinting at.  Nonetheless, the  error bars on data and fit leave room for the  $G_E^s$ prediction of most models,  
and the CQS $G_M^s$ value.  The global $Q^2$ dependence of Eqs.~(\ref{eq:f1forkel}) and (\ref{eq:f2forkel}), combined with the CQS vector strangeness values  seems consistent with the data for $G_E^s+\eta G_M^s$.  

\begin{table}[h]
\centering
\begin{tabular}{llll}
\hline
Model  & Ref.  &  $\mu_s(\mu_N)$ & $r_s^2$(fm$^{2}$)\\
\hline
$VMD$ &  \cite{jaffe89} &  -0.31 &  0.16 \\
$K\Lambda$ &  \cite{musolf94} &  -0.35 & -0.007 \\
NJL  & \cite{kim95}  & -0.45 &  -0.17 \\
CQS (K)  & \cite{silva01}  & 0.115  & -0.095 \\ 
\hline
\end{tabular}
\caption{Predictions for $r_s^2$ and $\mu_s$ in some hadron models.}
\label{tab}
\end{table}

\section{Strangeness in quasielastic neutrino-nucleus scattering}\label{secres}

\begin{figure}[h]
%figuur2
  \begin{center}
    \includegraphics[width=14cm]{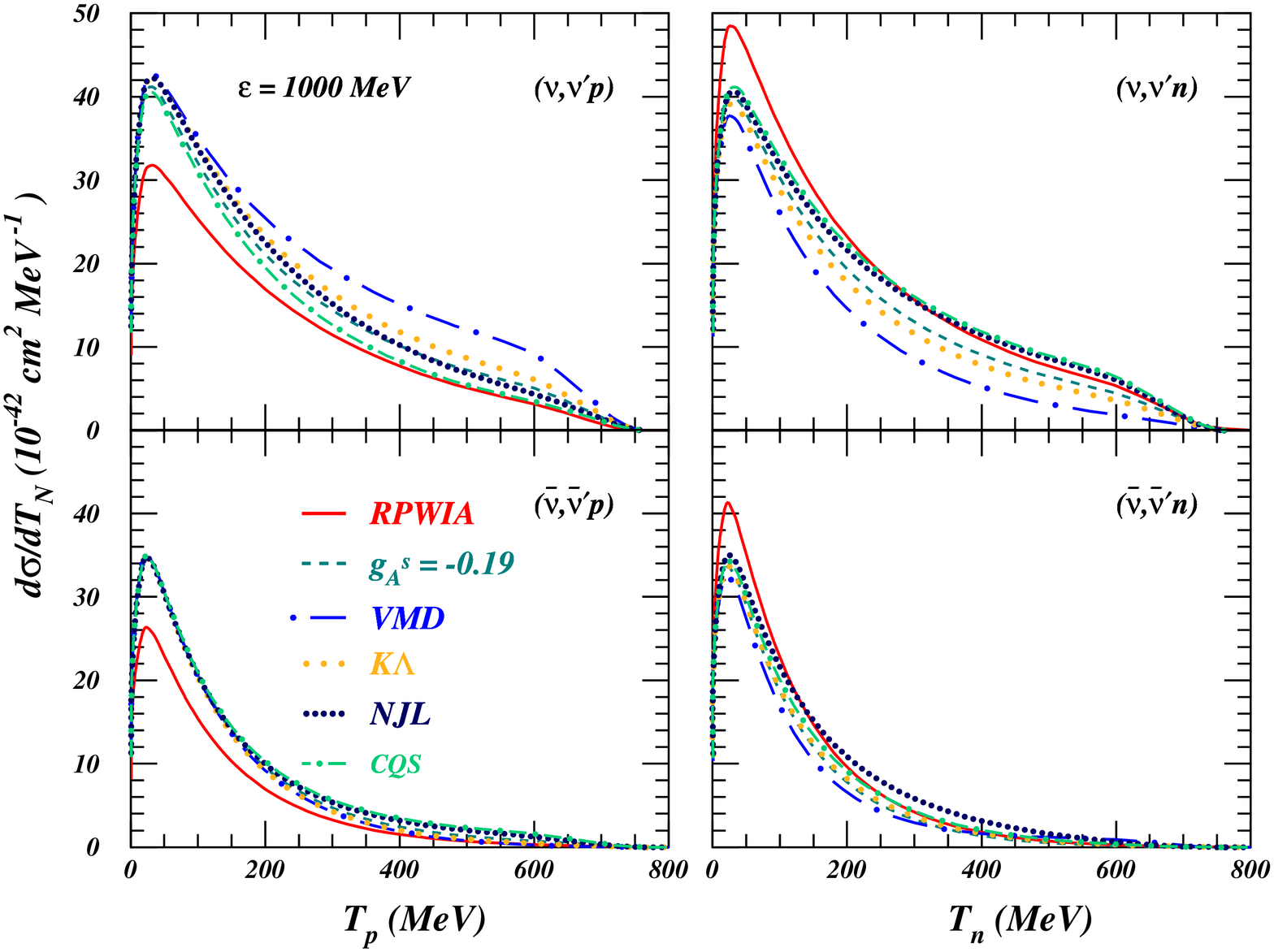}
   \caption{(Color online) Influence of sea quarks on the neutral current cross sections for
   $^{12}$C as a function of the outgoing nucleon kinetic energy $T_N$
   at an incoming energy of $\varepsilon \mbox{=} 1000$ MeV. The left (right) panel
   corresponds to proton (neutron) ejectiles. The solid curve
   represents the RPWIA results without strangeness. The other curves
   adopt $g_A^s \mbox{=} -0.19$ and correspond to different values for
   $r_s^2$ and $\mu_s$~: $(r_s^2 \mbox{=} 0, \mu_s \mbox{=} 0) $
   (short-dashed), VMD (long dot-dashed) \cite{jaffe89}, K$\Lambda$
   (long-dotted) \cite{musolf94}, NJL (short-dotted) \cite{kim95} and
   CQS(K) (short dot-dashed) \cite{silva01}.}
\label{fig:crossstrangep}
  \end{center} 
\end{figure}

The effect of a non-vanishing strange quark contribution to the axial and weak vector form factors on neutral current neutrino- and antineutrino-induced cross sections is shown in Fig.~\ref{fig:crossstrangep}.
The effect of the axial strangeness form factor $g_A^s$ is opposite for neutron and proton knockout reactions, causing a reduction of the cross section for neutron knockout reactions and a comparable enhancement  for reactions on a proton.  
For neutrino interactions, this behavior becomes more outspoken when adding the influence of the strange vector contributions.  This effect is most pronounced, albeit still rather small,  for the VMD and K$\Lambda$ models with positive or only very small strangeness radius. The positive $\mu_s$ value advocated by the CQS model and by  experimental data, tends to counterbalance the impact of $g_A^s$. The influence of the vector strangeness form factors in the NJL model is almost negligible.
For antineutrino-induced cross sections, the effect of weak vector strangeness contributions on the cross sections is marginal. The NJL model slightly enhances cross sections at large outgoing ejectile energies, in antineutrino reactions on a neutron.
The cross-section results of Fig.~\ref{fig:crossstrangep} clearly illustrate that when seeking strange quark effects in neutrino scattering, it is essential to differentiate between protons and neutrons.  The effects stemming from the strange sea-quarks tend to cancel when summing over proton and neutron contributions.
\begin{figure}[tbh]
%figuur3
  \begin{center}
\vspace*{12cm}
\special{hscale=82 vscale=82 hsize=1500 vsize=600
         hoffset=2 voffset=0 angle=0 psfile="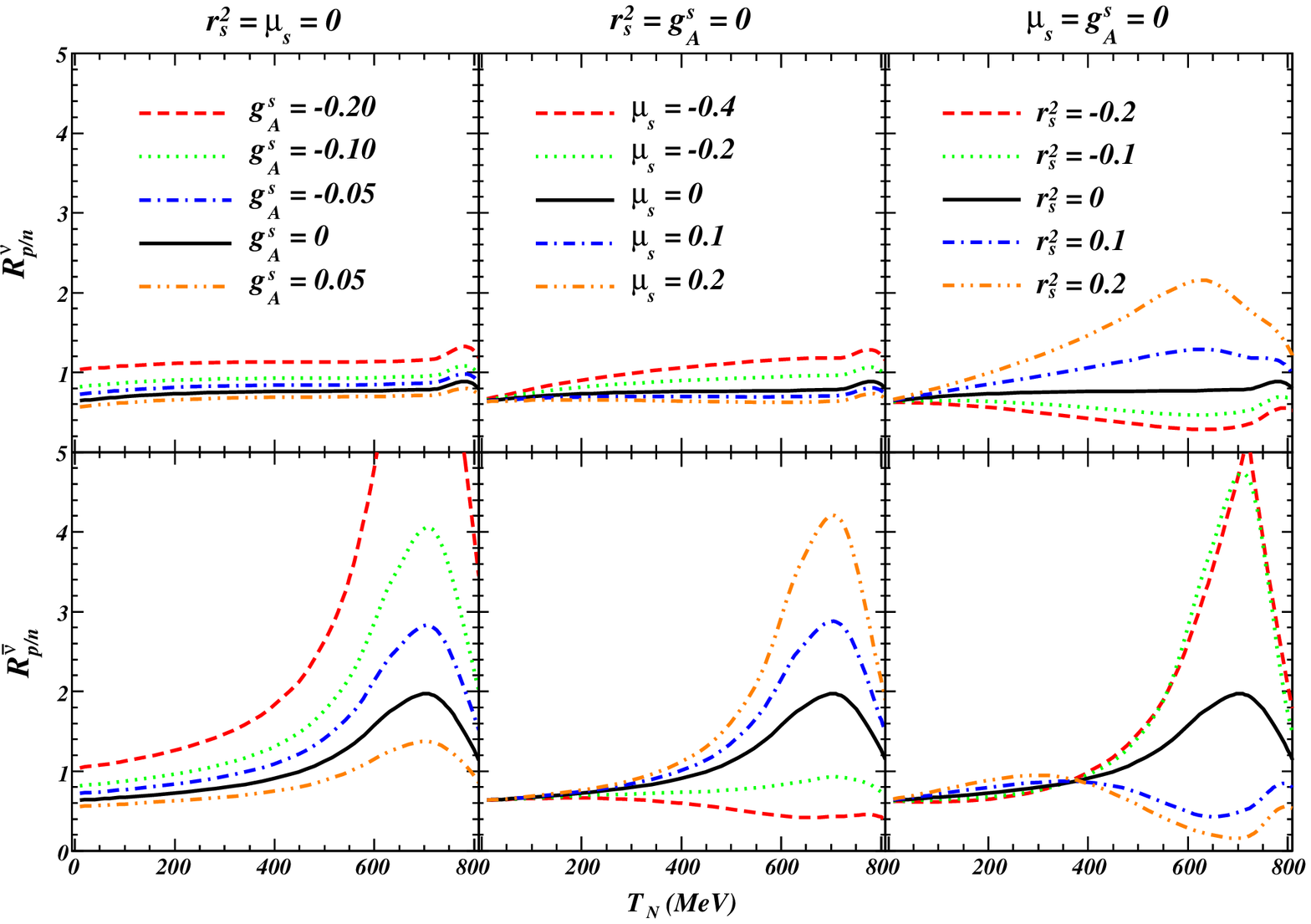"}
   \caption{(Color online) Effect of the strangeness parameters $g_A^s$, $\mu_s$ and $r_s^2$ on the proton-to-neutron neutrino (top panels) and antineutrino (bottom panels) cross-section ratios.  The calculations were performed for 1 GeV neutrinos  on a $^{12}$C target.  }
\label{Rpn}
  \end{center} 
\end{figure}

The  opposite influence of strangeness on      proton and neutron knockout processes  can be exploited when selecting the ratios that are most suitable for the extraction of strangeness information from neutrino scattering data.
The ratios
\begin{eqnarray}
R_{p/n}^{\nu}&=&\frac{\left(\frac{d\sigma}{dT_N}\right)^{NC}_{(\nu,p)}}{\left(\frac{d\sigma}{dT_N}\right)^{NC}_{(\nu,n)}},\\
R_{p/n}^{\overline{\nu}}&=&\frac{\left(\frac{d\sigma}{dT_N}\right)^{NC}_{(\overline{\nu},p)}}{\left(\frac{d\sigma}{dT_N}\right)^{NC}_{(\overline{\nu},n)}},
\end{eqnarray}
indeed use the differences in proton and neutron strangeness form factors to
maximize the strangeness asymmetry that can be obtained by combining cross sections. 
\begin{figure}[tb]
%figuur4
  \begin{center}
\vspace*{6.5cm}
\special{hscale=57 vscale=57 hsize=1500 vsize=600
         hoffset=-3 voffset=-10 angle=0 psfile="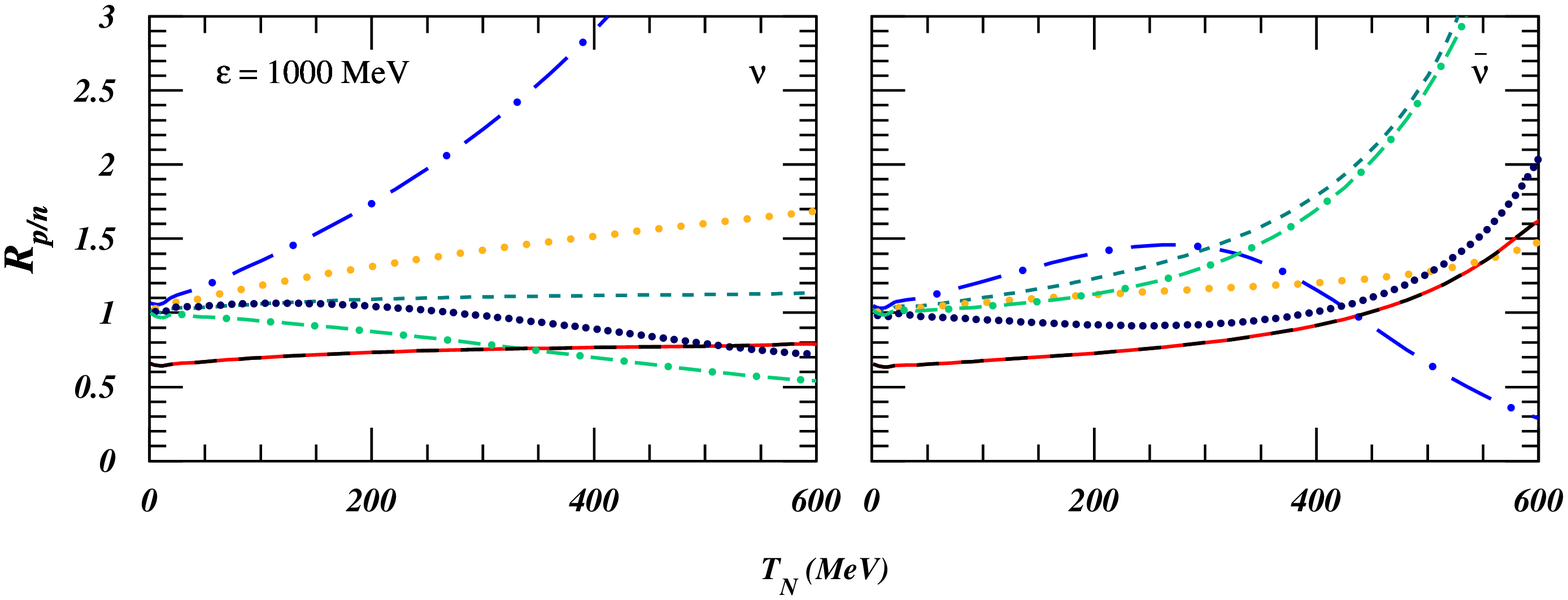"}
   \caption{(Color online) Ratio of proton-to-neutron neutral-current cross sections for
   quasielastic scattering on $^{12}$C. The left and right panels correspond to neutrino and antineutrino induced reactions respectively.    RPWIA results for vanishing strangeness (full line), and results including final-state interactions (long-dashed) coincide.
The line convention for the results including strangeness is the same as in Fig.~\ref{fig:crossstrangep} : $g_A^s \mbox{=} -0.19$ and $(r_s^2 \mbox{=} 0,\; \mu_s \mbox{=} 0) $
   (short-dashed), and RPWIA results for $g_A^s \mbox{=} -0.19$ and vector strangeness parameters $r_s^2$ and $\mu_s$ from the   VMD (long dot-dashed) \cite{jaffe89}, K$\Lambda$
   (long-dotted) \cite{musolf94}, NJL (short-dotted) \cite{kim95} and
   CQS(K) (short dot-dashed) \cite{silva01} models.}
\label{fig:rnurs}
  \end{center} 
\end{figure}

The left panels of Fig.~\ref{Rpn} display $R_{p/n}$ for both neutrino and antineutrino NC interactions on $^{12}$C, for different values of $g_A^s$ and vanishing vector strangeness form factors.  Clearly, this ratio is very sensitive to the value of the axial strangeness.  For $g_A^s=-0.20$ the neutrino-induced cross-section ratio $R_{p/n}^{\nu}$ nearly doubles compared to the values obtained for vanishing axial strangeness.  Whereas the ratio for neutrino-induced reactions remains remarkably constant as a function of $T_N$, its behavior is quite different for antineutrinos, showing ratios $R_{p/n}^{\overline{\nu}}$ that are growing steeply with increasing energy of the ejectile, where the cross sections become small.  This effect stems from the $Q^2$ evolution of the interplay  between axial and vector contributions to the response.
The middle panels of Fig.~\ref{Rpn} illustrate the sensitivity of $R_{p/n}^{\nu}$ and $R_{p/n}^{\overline{\nu}}$ to $\mu_s$ at vanishing $r_s^2$ and $g_A^s$.
Neutrino and antineutrino ratios show an opposite behavior, with growing $\mu_s$ values increasing the ratio for antineutrino induced reactions, and reducing neutrino-induced reaction ratios.  Again, the ratio is much more dependent on the ejectile kinetic energy in antineutrino reactions than in neutrino-induced cross-sections.  For sufficiently high $T_N$, antineutrino ratios are depending more strongly on the strangeness parameters than ratios of neutrino cross sections are.    Most striking is the large influence of $r_s^2$ on $R^{\nu}_{p/n}$, illustrated in the right panels of Fig.~\ref{Rpn}.  For antineutrino cross sections the influence of $r_s^2$ on the ratio is comparable to that of $g_A^s$ at large ejectile energies, and is smaller at low energies for the outgoing nucleon.

We examined the behavior of the proton-to-neutron ratio $R_{p/n}$  for the ($r_s^2$, $\mu_s$) parameterization of different hadron models, combined with an axial strangeness form factor $g_A^s=-0.19$.
Fig.~\ref{fig:rnurs} displays $R_{p/n}$ for both neutrino and
antineutrino NC interactions on $^{12}$C. The coinciding curves
representing the results with and without the effect of FSI, illustrate
the negligible influence of final-state interactions on these cross
section ratios \cite{lava,praet}.   This figure, as well as the left panel of Fig.~\ref{Rpn}, illustrates that  for $g_A^s=-0.19$, the $R_{p/n}$ ratios are enhanced by approximately 40\% in comparison with the $g_A^s=0$ situation.  
In neutrino cross sections, the VMD and K$\Lambda$ models tend to amplify
this enhancement, whereas the CQS predictions nearly cancel it.  The
behavior of the CQS model is especially interesting as this is the
only model that predicts a positive strangeness magnetic moment,
compatible with the PVES experiments. The influence of the NJL vector
strangeness is relatively small, what can be explained by  the results in the upper panels of Fig.~\ref{Rpn}, showing that the NJL $r_s^2$ and $\mu_s$ values have an opposite effect on the $R_{p/n}^{\nu}$ ratios. For antineutrinos, the vector
strangeness effects are more ambivalent.  With the NJL predictions for $\mu_s$ and $r_s^2$,  the axial strangeness effect is almost cancelled.  The influence of the CQS preditions on the other hand, is small.  
It is remarkable that, even when using the more traditional and relatively large 
$g_A^s=-0.19$ value, the axial vector effects tend to be overshadowed or canceled by those of the vector strangeness contributions, especially for the VMD and K$\Lambda$  predictions.

\begin{figure}[tbh]
%figuur5
  \begin{center}
\vspace*{11.5cm}
\special{hscale=82 vscale=82 hsize=1500 vsize=600
         hoffset=2 voffset=0 angle=0 psfile="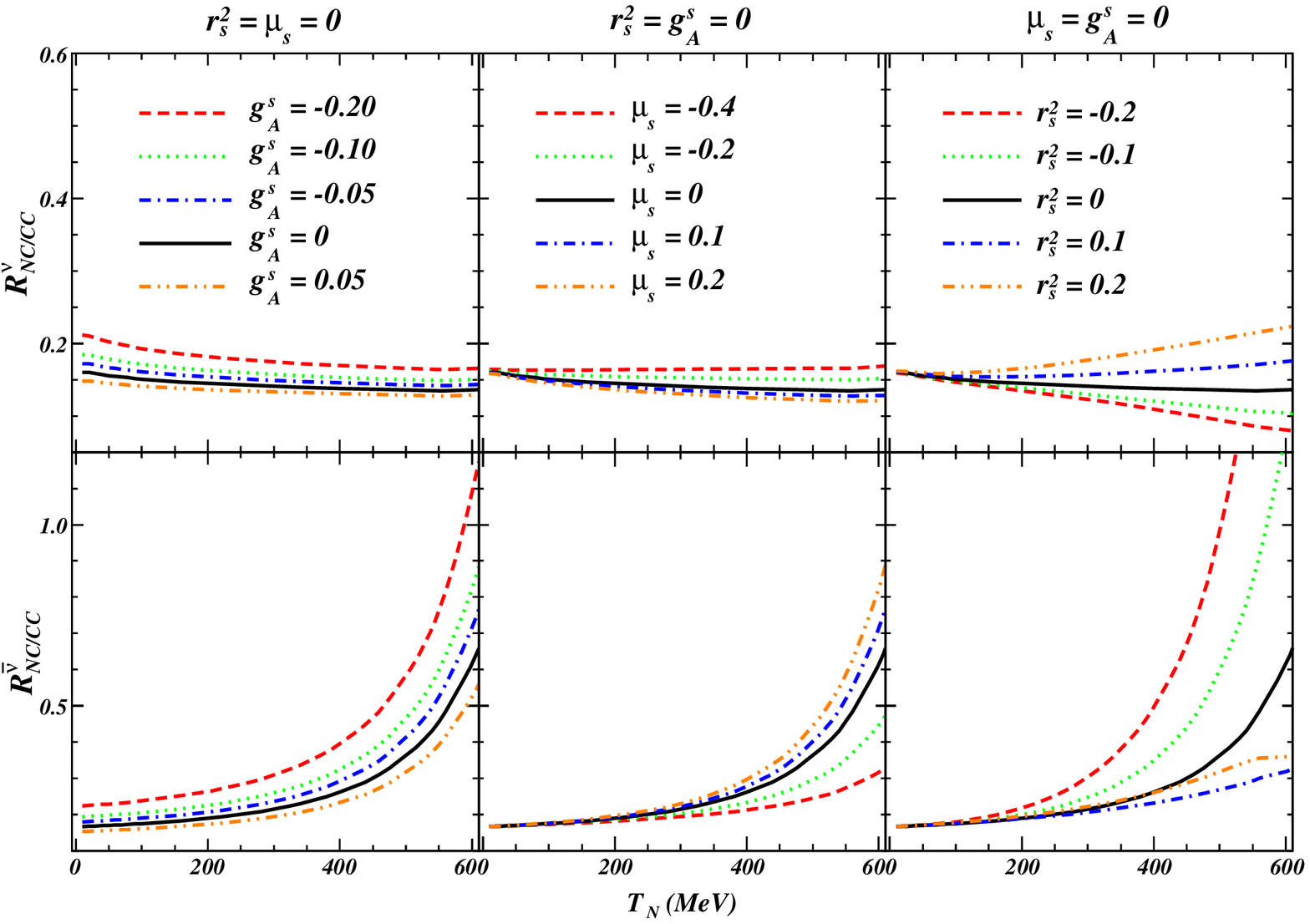"}
   \caption{(Color online) Effect of the axial and vector strangeness parameters $g_A^s$, $\mu_s$ and $ r_s^2$ on the neutral-to-charged current neutrino (top panels) and antineutrino (bottom panels) cross-section ratios as a function of the energy of the outgoing nucleon.  The calculations were performed for 1 GeV neutrinos  on a $^{12}$C target.  }
\label{Rnccc}
  \end{center} 
\end{figure}

\begin{figure}[bht]
%figuur6 
 \begin{center}
    \includegraphics[width=\linewidth]{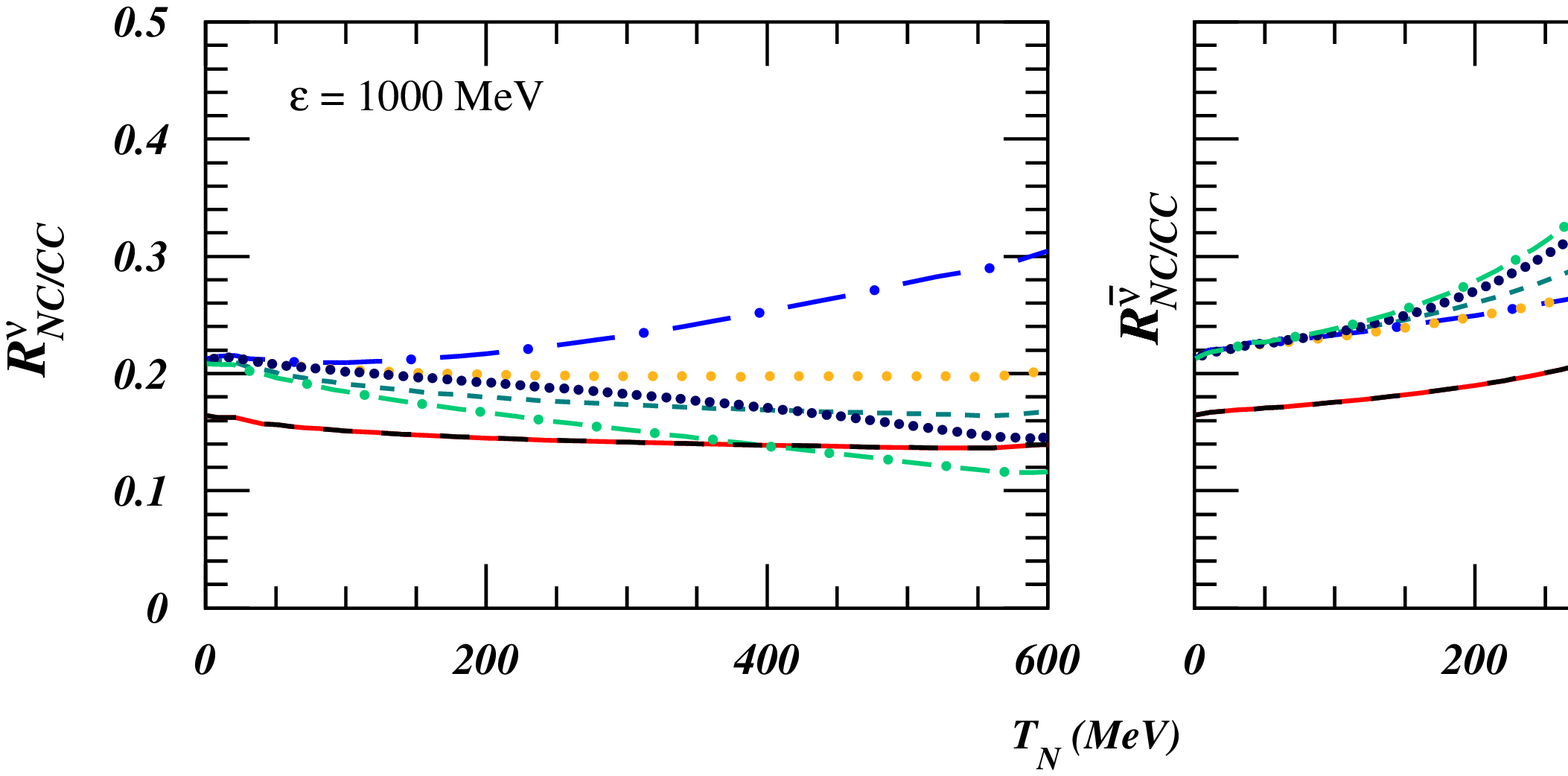}
   \caption{(Color online) Ratio of neutral-to-charged current cross sections for
   quasielastic scattering on $^{12}$C. The left (right) panels correspond
   to neutrinos (antineutrinos). RPWIA results  (full line), and results including final-state interactions (long-dashed) coincide. The line convention for the results including strangeness is the same as in Fig.~\ref{fig:crossstrangep}~:  RPWIA results for $g_A^s \mbox{=} -0.19$ and $(r_s^2 \mbox{=} 0,\, \mu_s \mbox{=} 0) $
   (short-dashed), results for $g_A^s \mbox{=} -0.19$ and VMD (long dot-dashed) \cite{jaffe89}, K$\Lambda$
   (long-dotted) \cite{musolf94}, NJL (short-dotted) \cite{kim95} and
   CQS(K) (short dot-dashed) \cite{silva01}.}
\label{fig:rnustrange}
  \end{center} 
\end{figure}

From an experimental point of view, the use of the proton-to-neutron ratio is tedious due to the difficulties inherent to neutron detection.  Therefore, it is often proposed to replace $R_{p/n}$ by the ratio  of neutral current over charged current (CC) cross sections
\begin{equation}
R^{\nu}_{NC/CC} = \frac{\sigma^{NC}(\nu p \rightarrow \nu p)}{\sigma^{CC}(\nu n \rightarrow \mu^- p)},
\end{equation}
\begin{equation}
R^{\overline{\nu}}_{NC/CC} = \frac{\sigma^{NC}(\overline{\nu} p \rightarrow \overline{\nu} p)}{\sigma^{CC}(\overline{\nu} p \rightarrow \mu^+ n)}.
\end{equation}
 Again, the difference between isoscalar and isovector contributions
 in  numerator and denominator is exploited to highlight the influence
 of strangeness.  Compared to $R_{p/n}$, the neutral-current proton-knockout cross section in the numerator remains unchanged, but the denominator is replaced by charged-current cross sections, that are more easily assessed experimentally. 
This reduces the strangeness effect compared to the one observed for the $R_{p/n}$ ratios. Indeed, instead of a ratio of quantities with a completely opposite behavior towards strangeness, the isovector denominator is blind to the strangeness content of the nucleon.

Figs.~\ref{Rnccc} and \ref{fig:rnustrange} provide a systematic study of the sensitivity of $R_{NC/CC}$  to   $g_A^s$, $\mu_s$ and $r_s^2$.
Qualitatively, the behavior of the  $R_{NC/CC}$ ratios is very similar to that of $R_{p/n}$.
Still, the cross-section ratios are smaller and the overall effect of  strangeness is reduced. This lowered sensitivity is most pronounced for $g_A^s$.  Except for the influence of $r_s^2$ on $R_{NC/CC}^{\overline{\nu}}$,  the $R_{NC/CC}$ ratios  are less sensitive to the strange vector form factors than  the proton-over-neutron ratios.

\begin{figure}[tbh]
%figuur7
  \begin{center}
\vspace*{11.5cm}
\special{hscale=82 vscale=82 hsize=1500 vsize=600
         hoffset=2 voffset=0 angle=0 psfile="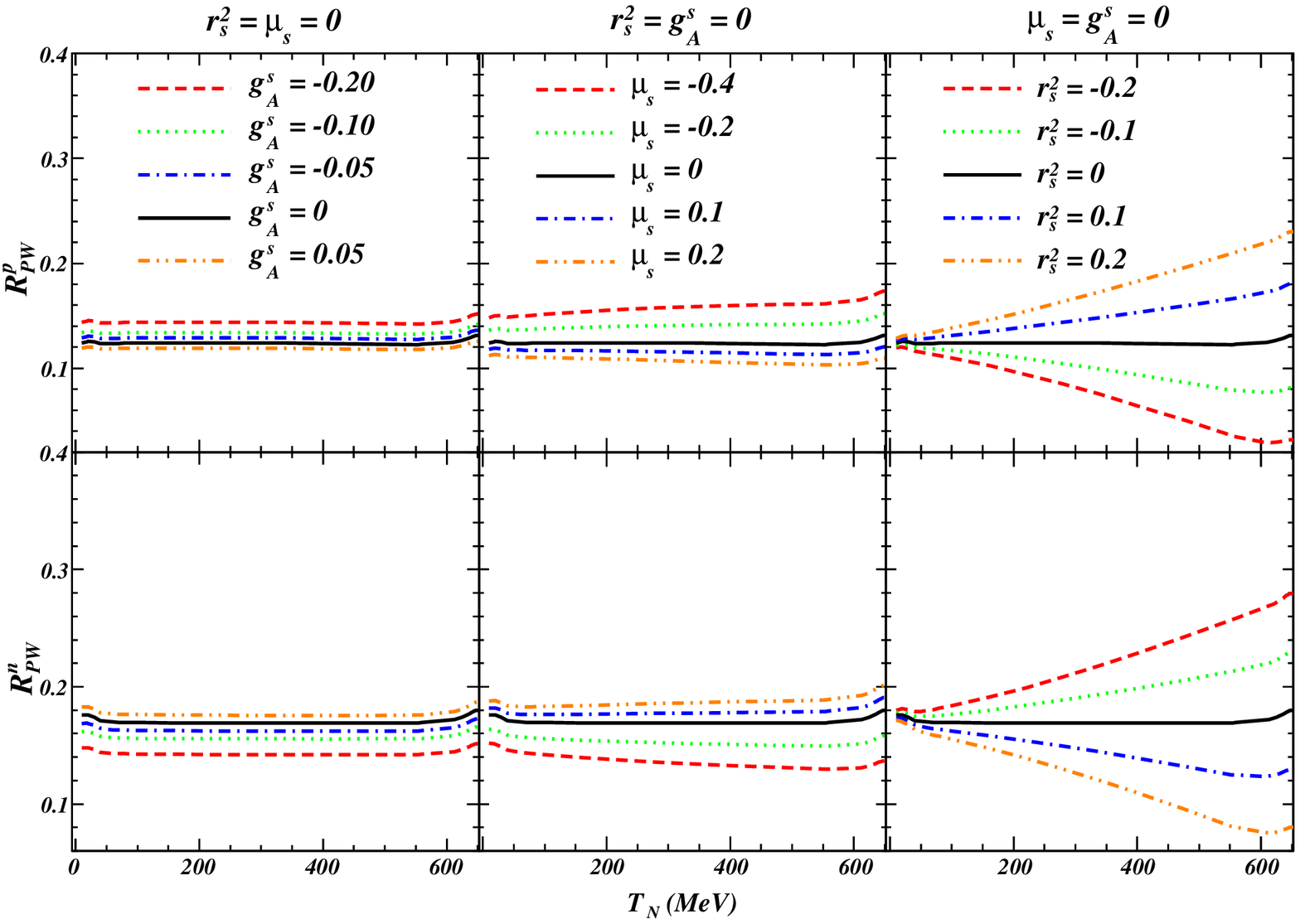"}
   \caption{(Color online) Effect of the strangeness parameters $g_A^s$, $\mu_s$ and $ r_s^2$ on the Paschos-Wolfenstein relation              for protons (top panels) and neutrons (bottom panels).  The calculations were performed for incoming neutrinos with an energy of 1 GeV  and a $^{12}$C target.  }
\label{RPW}
  \end{center} 
\end{figure}

In Refs.~\cite{praet,alberic} it was suggested that the ratios
\begin{eqnarray}
R^p_{PW}&=&\frac{\frac{d\sigma}{dT_N}(\nu p\rightarrow \nu p)-\frac{d\sigma}{dT_N}(\overline{\nu} p\rightarrow \overline{\nu} p)}{\frac{d\sigma}{dT_N}(\nu n\rightarrow \mu^- p)-\frac{d\sigma}{dT_N}(\overline{\nu} p\rightarrow{\mu}^+ n)},\\
R^n_{PW}&=&\frac{\frac{d\sigma}{dT_N}(\nu n\rightarrow \nu n)-\frac{d\sigma}{dT_N}(\overline{\nu} n\rightarrow \overline{\nu} n)}{\frac{d\sigma}{dT_N}(\nu n\rightarrow \mu^- p)-\frac{d\sigma}{dT_N}(\overline{\nu} p\rightarrow {\mu^+} n)},
\end{eqnarray}
can be useful in disentangling the strangeness properties of the nucleon.
In deep inelastic scattering the above  ratios are referred to as the Paschos-Wolfenstein relation (PW) and used to determine the weak mixing angle $\theta_W$.
Fig.~\ref{RPW} illustrates that they exhibit a sizable  sensitivity to the strangeness content of the nucleon.
Further, the influence of strangeness on  the PW relation for protons and neutrons is opposite.  Whereas the effect of $g^s_A$ on the $R_{PW}$ ratios is relatively modest, their sensitivity to $\mu_s$ and $ r_s^2$ is considerable.  For  $g^s_A$ and  $\mu_s$  the ratios are remarkably independent of the ejectile energy $T_N$.
The dependence of this ratio on $\sin^2\theta_W$ \cite{praet} might however complicate the extraction of strangeness information from PW-like relations.

\begin{figure}[tbh]
%figuur8
  \begin{center}
\vspace*{12cm}
\special{hscale=82 vscale=82 hsize=1500 vsize=600
         hoffset=2 voffset=0 angle=0 psfile="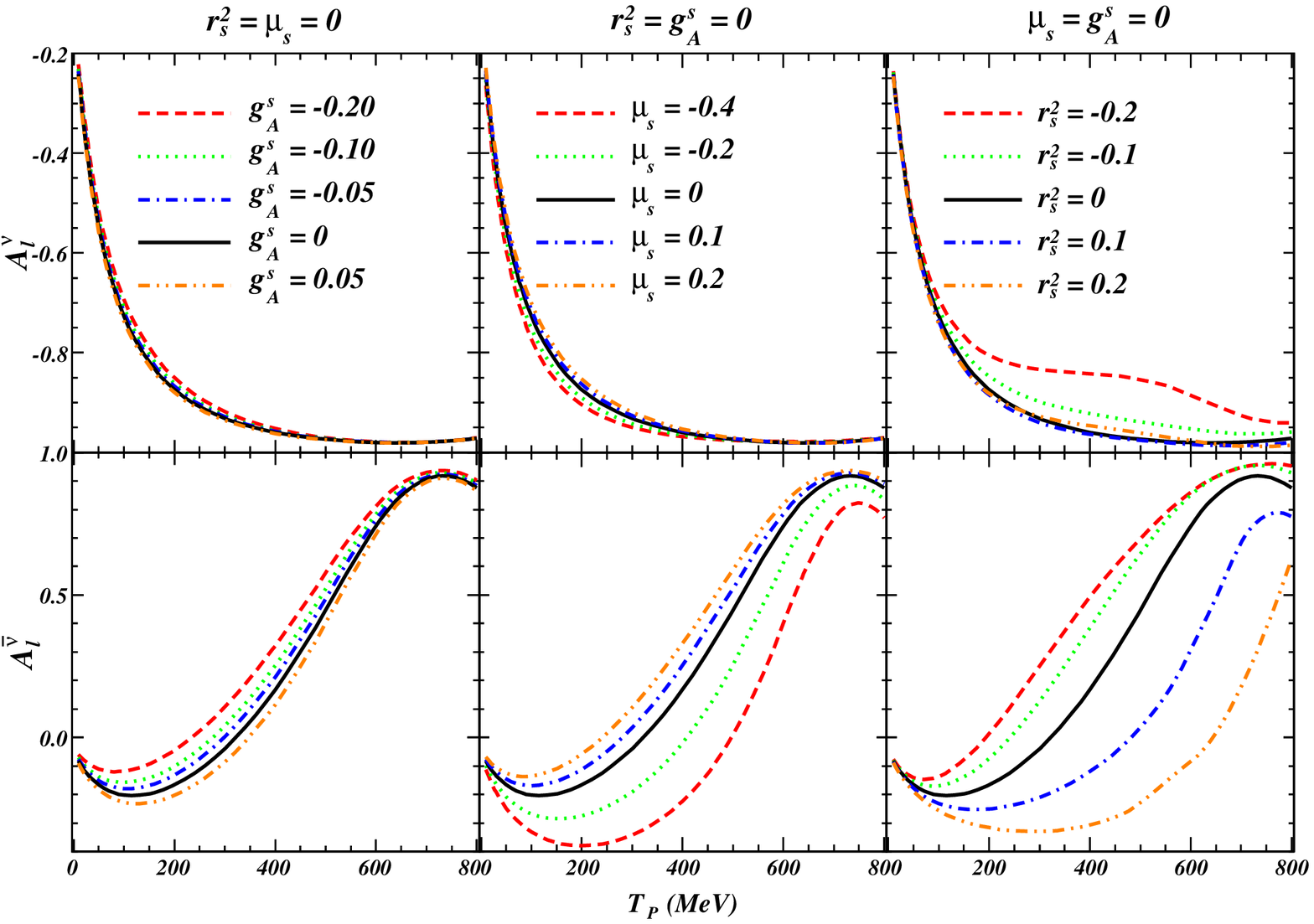"}
   \caption{(Color online) Effect of the axial and vector strangeness parameters $g_A^s$, $\mu_s$ and $r_s^2$ on the longitudinal helicity asymmetry for neutrinos (top panels) and antineutrinos (bottom panels).  The calculations were performed for 1 GeV neutrinos  on a $^{12}$C target.  }
\label{Rhela}
  \end{center} 
\end{figure}

Although its measurement would be extremely challenging, the longitudinal polarization asymmetry $A_l$ is an interesting quantity. It is defined as the difference  between cross sections for nucleon ejectiles with opposite helicities, 
 normalized to the total neutral-current nucleon-knockout cross section \cite{letter,lava}.  In terms of differential cross sections for proton knockout processes this results in
\begin{eqnarray}
A^{\nu}_l&=&\frac
{\frac{d\sigma}{dT_p}(\nu p\rightarrow \nu p,\,h_p=+1)-\frac{d\sigma}{dT_p}(\nu p\rightarrow \nu p,\,h_p=-1)}
{\frac{d\sigma}{dT_p}(\nu p\rightarrow \nu p,\,h_p=+1)+\frac{d\sigma}{dT_p}(\nu p\rightarrow \nu p,\,h_p=-1)}, \\
A^{\overline{\nu}}_l&=&\frac
{\frac{d\sigma}{dT_p}(\overline{\nu} p\rightarrow \overline{\nu} p,\,h_p=+1)-\frac{d\sigma}{dT_p}(\overline{\nu} p\rightarrow \overline{\nu} p,\,h_p=-1)}
{\frac{d\sigma}{dT_p}(\overline{\nu} p\rightarrow \overline{\nu} p,\,h_p=+1)+\frac{d\sigma}{dT_p}(\overline{\nu} p\rightarrow \overline{\nu} p,\,h_p=-1)}.
\end{eqnarray}
The peculiar structure of these quantities, combined with the fact that the neutrino projectiles are fully polarized, allows one to select particular contributions to  the hadron response.  Fig.~\ref{Rhela} shows  the dependence of $A^{\nu}_l$ on strangeness effects to be almost negligible. Its antineutrino counterpart $A^{\overline{\nu}}_l$ on the other hand, is extremely sensitive to strangeness mechanisms. $A^{\overline{\nu}}_l$ is  unmatched as a filter for axial strangeness effects,  and  exhibits an even stronger  sensitivity to the vector strangeness form factors.

\section{FINeSSE}\label{secfin}

\begin{figure}[tbh]
%figuur9
  \begin{center}
\vspace*{6.5cm}
\special{hscale=44 vscale=44 hsize=1500 vsize=600
         hoffset=100 voffset=0 angle=0 psfile="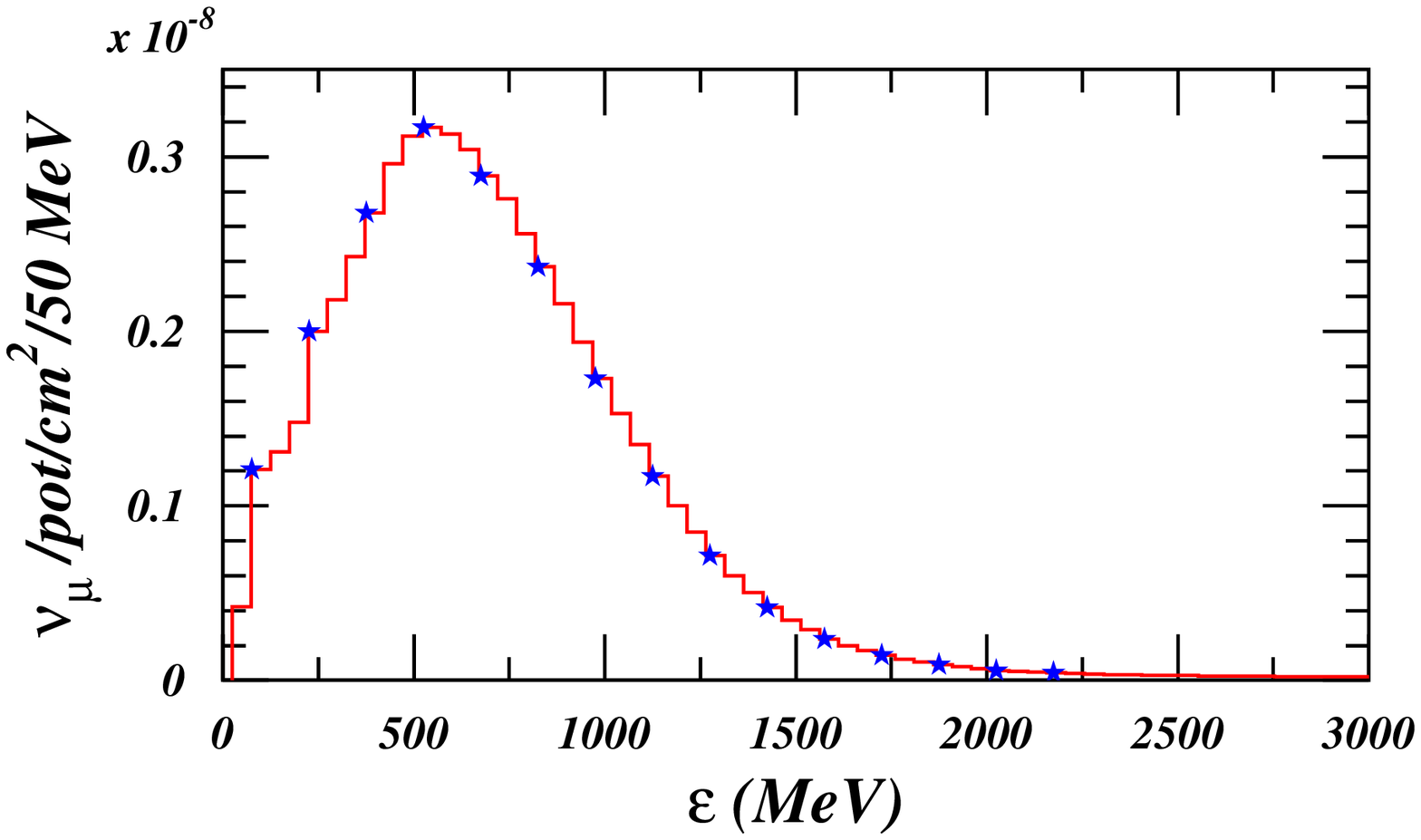"}

   \caption{(Color online) A typical FINeSSE flux on the FNAL
   booster neutrino beamline for neutrinos  \cite{patepriv}. The average beam energy
   corresponds to $\langle \varepsilon \rangle \approx 700$ MeV. The
   markers indicate the energies for which calculations were
   performed.}
\label{fig:aflux}
  \end{center} 
\end{figure}

As actual neutrino experiments will employ neutrinos of a quite broad energy distribution rather than a monochromatic neutrino beam, we repeated the  analysis of $R^{\nu}_{NC/CC}$, folding the cross sections with an appropriate experimental neutrino energy spectrum.
FINeSSE aims at probing the  strangeness content of the nucleon at low $Q^2$, using the neutral-to-charged current cross-section ratio.  As shown in Fig.~\ref{fig:aflux}, the proposed neutrino beam has an average energy of 700 MeV, with tails up to 2 GeV.
The flux-averaged differential cross section is defined as
\begin{equation}
\left\langle \frac{d\sigma}{dT_N} \right\rangle =
\frac{\int_{\varepsilon_{min}}^{\varepsilon_{max}}\Phi(\varepsilon)\frac{d\sigma}{dT_N}(\varepsilon)d\varepsilon
}{\int_{\varepsilon_{min}}^{\varepsilon_{max}}\Phi(\varepsilon)d\varepsilon
}. \; 
\end{equation}
with $\Phi(\varepsilon)$ the typical FINeSSE (anti)neutrino spectrum.

\begin{figure}[tbh]
%figuur10
  \begin{center}
    \includegraphics[width=\linewidth]{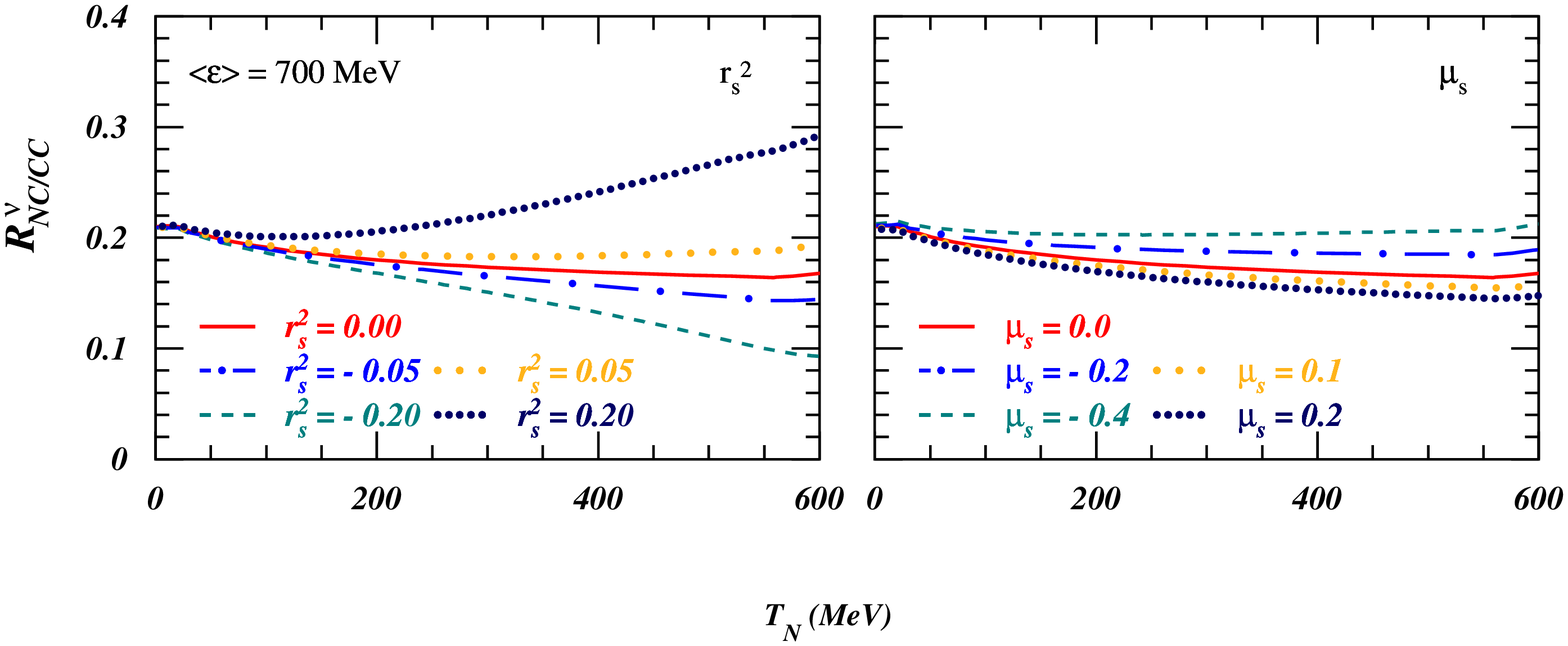}
   \caption{(Color online) Neutral-to-charged current ratio of flux-averaged  cross sections for
   quasielastic scattering on $^{12}$C. All calculations are performed within the RPWIA,  with $g_A^s \mbox{=} -0.19 $. The left (right) panel
   illustrates the predicted effect of varying the strangeness radius
   (magnetic moment), putting $\mu_s=0$ ($r_s^2=0$).}
\label{fig:rnursfiness}
  \end{center} 
\end{figure}

Fig.~\ref{fig:rnursfiness} shows the influence of the folding on the behavior of the ratios.  The axial strangeness form factor was kept fixed at $g_A^s=-0.19$, the vector strangeness parameters were varied within the range indicated by the models in Table \ref{tab}.  Comparing these results with the ones of Fig.~\ref{Rnccc}, illustrates that the folding only slightly modulates the picture. Qualitatively, the strangeness influence on the cross section ratio $R_{NC/CC}^{\nu}$ remains the same within the  energy ranges relevant for the FINeSSE experiment.

\section{Comparing ratios}\label{conclu}

\begin{figure}[tbh]
%figuur11
\vspace*{5.5cm}
  \begin{center}
\special{hscale=82 vscale=82 hsize=1500 vsize=600
         hoffset=2 voffset=0 angle=0 psfile="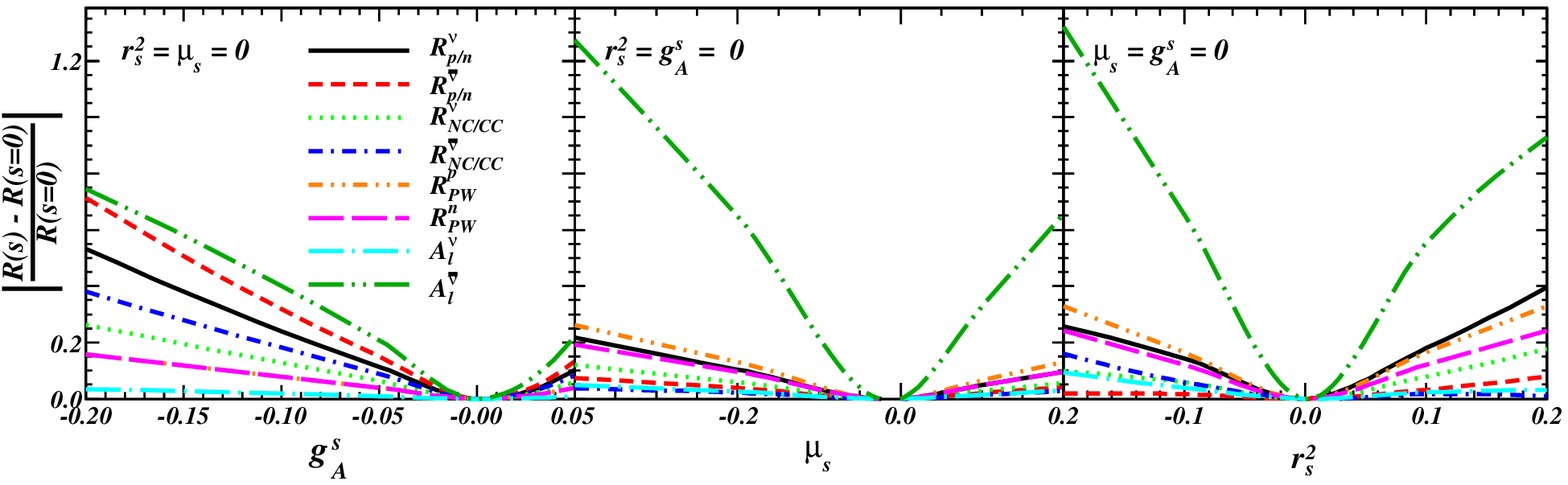"}
   \caption{(Color online) Comparison between the strangeness influence on various ratios of total cross-sections in terms of the relative sensitivity $\left|\frac{R(s=0)-R(s)}{R(s=0)}\right|$, as a function of the strangeness form factors $g_A^s$, $\mu_s$ and $ r_s^2$ for 1 GeV neutrino scattering off $^{12}$C.}
\label{vergelijkingRatio}
  \end{center} 
\end{figure}
\begin{figure}[tbh]
%figuur12
\vspace*{9.0cm}
  \begin{center}
\special{hscale=82 vscale=82 hsize=1500 vsize=600
         hoffset=2 voffset=0 angle=0 psfile="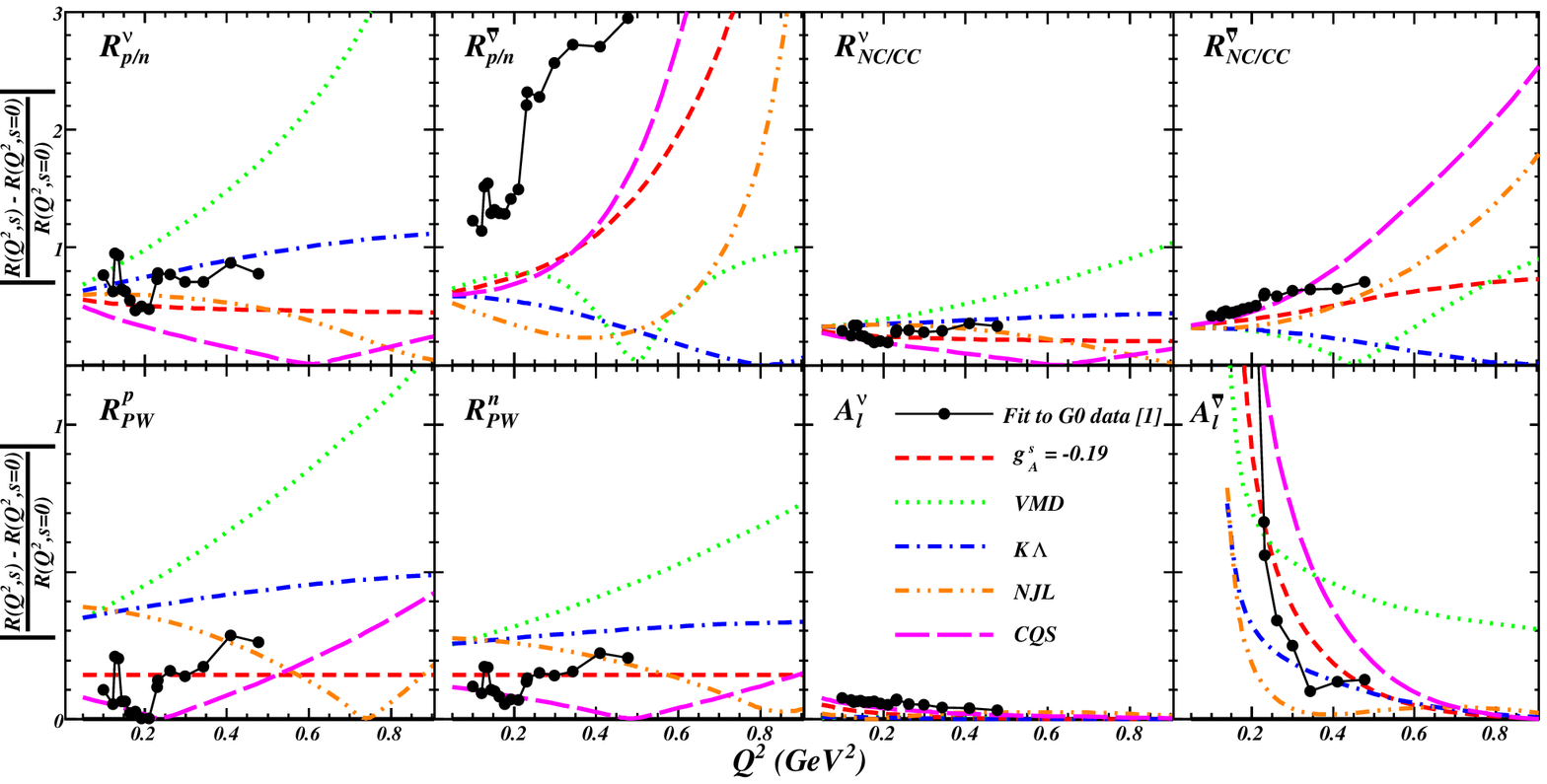"}
   \caption{(Color online) $Q^2$ dependence of the strangeness influence on cross-section ratios in terms of the relative strangeness sensitivity $\left|\frac{R(Q^2,s=0)-R(Q^2,s)}{R(Q^2,s=0)}\right|$, for $g_A^s=-0.19$, combined with the vector strangeness form factors provided by the hadron models VMC, K$\Lambda$, NJL, and CQS.  The incoming neutrino energy is 650 MeV, very close to the average neutrino energies proposed for FINeSSE, $^{12}$C is the target nucleus.}
\label{Q2dependDubbelRatio}
  \end{center} 
\end{figure}

In Figs.~\ref{vergelijkingRatio} and ~\ref{Q2dependDubbelRatio}, we summarize the results observed in Figs.~\ref{Rpn}-\ref{Rhela} and \ref{fig:rnursfiness}.  The figures compare the sensitivity of the different ratios to the strangeness parameters $g_A^s$, $\mu_s$ and $ r_s^2$, quantified by 
\begin{equation}
\left|\frac{R(s=0)-R(s)}{R(s=0)}\right|,
\end{equation}
with $R$ representing the ratios $R_{p/n}$, $R_{NC/CC}$, $R_{PW}$ or the helicity asymmetry $A_l$.
Fig.~\ref{vergelijkingRatio} highlights the effect of the strangeness form factors on ratios of integrated cross sections and compares the sensitivity of these ratios to the different strangeness parameters.  
Clearly, the antineutrino helicity asymmetry $A_l^{\overline{\nu}}$ has no equal when it comes to probing strangeness effects. It is the sole quantity that is more sensitive to the vector than to the axial strangeness parameters.
Nonetheless, only $R_{p/n}^{\overline\nu}$ can compete with $A_l^{\overline{\nu}}$ in its sensitivity to $g_A^s$. For most ratios, the antineutrino version exhibits a stronger strangeness sensitivity than the ratio constructed using neutrino-induced cross sections.
The sensitivity of some ratios strongly depends on
the sign of the strangeness parameters.  $R_{p/n}^{\nu}$ is mainly
sensitive to  positive values of $r_s^2$, while
the opposite is the case for $R_{PW}^p$. $R_{p/n}^{\overline{\nu}}$
and $R_{NC/CC}^{\overline{\nu}}$ offer good perspectives  in obtaining
$g_A^s$ information, and are not affected too much by the influence of $r_s^2$ and $\mu_s$.  The Paschos-Wolfenstein relation on the other hand, is most sensitive to the vector strange form factors while its sensitivity to $g_A^s$ is rather marginal.
  
Fig.~\ref{Q2dependDubbelRatio} illustrates the strangeness sensitivity of the cross-section ratios as a function of their $Q^2$ dependence. For measurements at low $Q^2$, the most suitable tools for the extraction of vector strangeness information are the Paschos-Wolfenstein relation for protons and the longitudinal helicity asymmetry for antineutrinos.
Even for models with  small strange vector form factors as CQS and K$\Lambda$,  the effect on the cross-section ratios can be  huge at relatively small values of the four-momentum transfer.
The strong $Q^2$ fluctuations in some of the curves, are due to the competition between the effects
 of $g_A^s$, $G_E^s(Q^2)$ growing fast at low $Q^2$, and  $G_M^s(Q^2)$ losing influence at higher $Q^2$.   The large influence of the vector strangeness values advocated by the data reflects the relatively large absolute values of the fitted values for $G_E^s(Q^2)$ and $G_M^s(Q^2)$ from Ref.~\cite{Liu}. 
The  $g_A^s$ dependence of the $R_{NC/CC}$ ratios
is relatively free of vector strangeness effects up to relatively high momentum transfers.

In conclusion, we provided a systematic overview of the sensitivity of neutrino cross-section ratios to the strange quark content of the nucleon. We presented a variety of (anti)neutrino cross-section ratios and compared the influence of axial as well as vector strangeness in terms of ejectile energies and $Q^2$ values. The longitudinal helicity asymmetry for antineutrinos is most sensitive to strangeness effects.  In general, antineutrino-induced processes exhibit a more outspoken strangeness sensitivity than their neutrino counterparts.
The overall sensitivity of $R_{NC/CC}$ ratios to strangeness effects is considerably smaller than that of $R_{p/n}$, but at small $Q^2$, the strangeness contributions to $R_{NC/CC}$ are more strongly dominated by the axial channels.

Although neutrino scattering is usually regarded as an excellent lever for  extracting  information about the axial strangeness, strange vector form factors have a remarkably strong influence on these ratios.
Whereas in PVES the tininess of the axial strangeness effects  impedes the determination of $g_A^s$, in neutrino scattering a thorough understanding of vector strangeness effects is a prerequisite for extracting information on the axial strangeness. Hence a combined  analysis of  parity-violating electron scattering and neutrino-induced processes would  offer the best prospects for a thorough understanding of the influence of the nucleon's strange quark sea on electroweak processes.

\acknowledgments
This work was supported by the Fund for Scientific Research (FWO), Flanders
and the Research Council  of Ghent University.

\end{document}